# Can Global, Extended, and Repeated Ransomware Attacks Overcome the User's Status Quo Bias and Cause a Switch of System?

Alex Zarifis, University of Nicosia, Cyprus

https://orcid.org/0000-0003-3103-4601

Xusen Cheng, Renmin University of China, China

Uchitha Jayawickrama, Loughborough University, UK

https://orcid.org/0000-0002-7159-6759

Simone Corsi, Loughborough University, UK

**ABSTRACT**

Ransomware (RW) attack effectiveness has increased causing far-reaching consequences that are not fully understood. The ability to disrupt core services, the global reach, extended duration, and the repetition has increased their ability to harm organizations. One aspect that needs to be understood better is the effect on the user. The user in the current environment is exposed to new technologies that might be adopted, but there are also habits of using existing systems. The habits have developed over time with trust increasing in the organization in contact directly and the institutions supporting it. This research explores whether the global, extended, and repeated RW attacks reduce the trust and inertia sufficiently to change long-held habits in using information systems. The model tested measures the effect of the RW attack on the e-commerce status quo to evaluate if it is significant enough to overcome the user's resistance to change.

## INTRODUCTION

In the story of David versus Goliath, an underdog manages to beat a much larger and stronger opponent. This metaphor can be used to describe the ransomware (RW) attacks. They may have limited resources like David while the organizations being targeted and the institutions supporting them are often like Goliath with extensive resources. We would like to believe, in this case, that the large organizations and institutions will emerge victorious by limiting the harm inflicted on them and their users. Is this, however, the case? The user has come to expect a reliable service from the train operators, airports and other services and products they use with minimal delay or downtime. Most users also experience secure transactions, secure storage and responsible use of personal information.

Examples of core service failure such as extended periods without access to services are rare and are usually limited to an economic crisis and failing organizations (Mansfield-Devine, 2020). Most users have also not experienced breaches of security that would reveal their personal information (Simoiu, Symantec, Bonneau, & Goel, 2019). This has built a trust in the institutions, organizations and the way personal information is handled. It has also created an e-loyalty (Carter, Wright, Thatcher, & Klein, 2014) expressed as an inertia and habit of the user in favour of the current systems used in e-commerce (Polites & Karahanna, 2012). The user however, is now facing the new phenomenon of global, extended and repeated RW attacks. While many users are directly affected by these attacks the reports in the media, social media and word of mouth serve to further magnify the impact. This may cause a momentary, or more extended, erosion of trust in the organizations they are directly in contact with and the institutions that support them. These attacks may also influence the user to such a degree that they overcome the inertia they have in favour of existing systems.

RW attacks use a malware to encrypt files on a computer and request a monetary amount, usually in Bitcoin, for the files to be unencrypted and made available for use again (Mercaldo, Nardone, & Santone, 2016). Ransomware attacks cost approximately 45 billion dollars in 2018 (Online Trust Alliance 2019). While each attack may have some variation in how the computer is infected, what files are encrypted and how the encryption is reversed, they are similar in their approach (Kharraz, Robertson, Balzarotti, Bilge, & Kirda, 2015). This form of malware is not new but its ability to disrupt an organization's core services repeatedly and for a prolonged period has increased. The effectiveness has increased because a combination of technologies and circumstances, are more favourable now than ten or fifteen years ago. For example, technologies such as digital currencies and circumstances such as outdated, unsupported operating systems have enabled and amplified attacks (Kshetri & Voas, 2017).

Recent RW attacks such as WannaCry, Petya, NotPetya, exPetr, Bad Rabbit, Sodinokibi-REvil (Simoiu et al., 2019; Yaqoob et al., 2017) are critical incidents that may have had an impact on the user and the willingness to engage in e-commerce as they did before. Since the start of the century business to consumer e-commerce has expanded with more people adopting it and existing users utilizing it more regularly. These repeated attacks may erode trust and loyalty. The user may stop engaging with the online vendor they had a habit of using if that vendor is attacked. A switch might be made to an online vendor that has not been attacked or an offline vendor less dependent on information systems. The user may switch to a new solution completely or partially. For example, the user may continue to use the same vendor but limit the value exchanged or the personal information shared. Lastly, the decision may be made to abstain from the exchange of value that was intended to be made. Improving the understanding of this phenomenon on the e-commerce user, will enable remedial action to be taken before, during and after an attack. Therefore:

*The aim of this research is to identify the factors that influence the user's decision to stop using an organization's system because of a RW attack.*

This research combines studies on inertia and resistance to switching systems (Polites & Karahanna, 2012) with a more comprehensive set of variables that cover the current e-commerce status quo. Personal information disclosure is included along with inertia and trust as it is now integral to e-commerce functioning effectively. The model developed captures the cumulative effect of this form of attack and evaluates if it is sufficiently harmful to overcome the e-loyalty and inertia built over time.

The implications of this research are both theoretic and practical. The theoretic contribution is highlighting the importance of this issue to IS theory, linking the RW literature to user inertia in IS and developing a model. There are three practical implications. Firstly, by better understanding the impact on the user it may be possible to have a new strategy to reduce the effectiveness of RW attacks. Secondly, processes can be created to manage such disasters as they are happening and maintain a positive relationship with the user. Lastly, the organizations can develop a buffer of goodwill and e-loyalty that would absorb the negative impact on the user from an attack and stop them reaching the point where they decide to switch system.

## BACKGROUND

### Ransomware Attacks

New RW attacks are emerging regularly as the attackers try to overcome the latest software updates and security solutions. Google identified 34 variations of RW in 2017 (Ramsey, 2017) and the number has since increased with new attacks like STOP(DJVU) in 2019. Despite their large number, they share similar characteristics and most new attacks are evolutions of previous ones. RW attacks are a combination of malware and extortion. The attack has a number of steps, as illustrated in figure 1: Firstly infection, secondly encryption and possibly deletion and thirdly reversal if the ransom is paid.

**Figure 1. The steps of a typical ransomware attack**

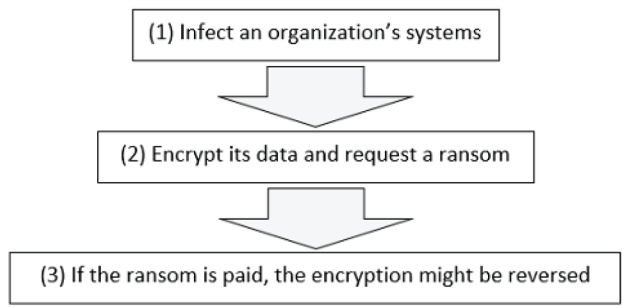

The infection stage, for the more primitive RW attacks, typical before 2016, usually required an action by the victim (Chang & Seow, 2019). This could include an attack known as a 'web drive-by' where a user downloads and runs a file that appeared to be a software update while visiting a legitimate cite (Erridge, 2016). There are also more targeted socially engineered infections that send senior staff an email with an attachment that appears to be a compressed document file. More recent attacks such as WannaCry and NotPetya, possibly with the exception of Bad Rabbit (Mamedov, Sinitsyn, & Ivanov, 2017), did not require an action by the victim. For the RW attacks where no action from the victim is needed, the malware may be able to move from one computer to another across an organization's network. This is a major factor in RW of this type spreading quickly. In these cases, the vulnerability comes primarily from the platform provider, rather than the user. Therefore, these platform providers that offer software and hardware, play an important role. It is not just the organization that comes into direct contact with the user, that has an influence. While the user may not initially know who these institutions are, this information can emerge during the attack.

Once a computer is infected, certain files are encrypted and a ransom request appears on the screens of the infected systems. The more sophisticated and extensive the encryption is, the harder it is to reverse it without paying the ransom. The malware used can randomly encrypt some files or target specific files that are more valuable. Files can also be targeted to make the recovery harder such as computer logs of the event. The RW message is designed to cause fear and panic in the victim. It usually includes a large countdown, many different fonts and flashing text. The ransom is typically requested in Bitcoin, usually half a bitcoin. This payment method is chosen so that it is harder to identify the attacker (Zimba, Wang, & Mulenga, 2019). Sometimes, a further threat that the ransom will be increased, is included. The ransom request can be forced on to all monitors across an organizations network including information screens in public places such as retail stores, train stations and hospitals. This puts pressure on the users or consumers which will add additional pressure on the organization to pay the ransom.

For the victim to make the ransom payment, a digital currency such as Bitcoin must be acquired first. This is a process most victims may not be familiar with. A digital currency is also a technology, with certain software and knowledge necessary to use it. Therefore, it can be considered that the victim must first adopt a new technology before paying the ransom. In some cases, when the ransom is payed the attacker reverses the encryption using a decryption key or 'kill switch' and removes the malware as the ransom request promised. However, in other cases, the attacker takes no action to reverse the harm done. The encryption may not be reversed out of choice or because the attacker does not have the ability to reverse it. This highlights that an attack can either be controlled across all the stages by the attacker, or it can be caused by self-propagating malware that is no longer controlled by anyone (Simoiu et al., 2019).

The number of RW attacks has increased from one in 2012 to 193 in 2016 (Kshetri & Voas, 2017) and 184 million in 2018 (SonicWall, 2019). The first attack that caused core service failures across many countries and could be considered global, was WannaCry in May 2017 affecting over 200000 systems in over 150 countries (Young & Yung, 2017). This was followed by the NotPetya attack in June, 2017. This was a more advanced version of Petya that had started appearing on a smaller scale 2016. In October of the same year, the Bad Rabbit attack emerged, initially targeting government information systems. In 2019 the new Sodinokibi-REvil attack caused large organizations to stop operating (Mansfield-Devine, 2020). In the future, attacks may increase further in their scale and ability to disrupt by targeting connected devices. As the Internet of Things (IoT) becomes more widespread, our dependence on connected devices will increase and the number of vulnerable targets will increase also (Yaqoob et al., 2017).

## E-Commerce Status Quo And Inertia

### Inertia and E-Loyalty

Many of us have used B2C e-commerce for several years now. In such cases the relationship between the user and the organization has gone beyond technology adoption to a habit and e-loyalty (Carter et al., 2014; Gefen, 2003; Limayem, Hirt, & Cheung, 2007). Continued behavior that follows the status quo instead of a superior alternative, increases inertia (Polites & Karahanna, 2012). The information systems and the e-service provided to the user in the current e-commerce environment, satisfies many of them and creates this inertia. As the technology and business processes supporting e-commerce improve over time, the user is regularly presented with a better service, so the satisfaction is maintained or increased. Therefore, there can be an accumulation of goodwill over the years. The inertia is compounded by switching costs. Engaging in e-commerce with a specific organization or platform requires an investment of time to complete processes such as registering and sharing of personal details. In addition to the inevitable cost of changing system, switching barriers are also introduced intentionally to make the change harder (Ghazali, Nguyen, Mutum, & Mohd-Any, 2016). Consequently, there can be a significant resistance from the user to switching systems. This resistance to switching can be overcome by a critical incident that is sufficiently influential.

User and consumer switch resistance has been evaluated in relation to other critical incidents apart from RW attacks (Polites & Karahanna, 2012). It has not been sufficiently evaluated for this type of incident, particularly when it is repeated. Literature suggests the user's resistance to changing service is different for different types of incidents. It is therefore useful to evaluate the effect of this type of incident on the user. For a user to switch system they need an alternative (Ghazali et al., 2016). If an incident has made them aware of weaknesses in the current system they use, an alternative may be more appealing as long as it does not appear to have the same weaknesses. While the attacks are extensive, so far, they have not impacted all the systems and all the organizations in one sector of the economy. Therefore, there are organizations and systems that have not been attacked, that may appear to be an appealing alternative for the user to switch to. The switch may not be complete and instant. There can be a transitionary period taking some time (Zeithaml, Berry, & Parasuraman, 1996). As the goodwill, habit and e-loyalty have built up over time with a series of successful exchanges of value, it may take a series of negative events to overcome the inertia.

E-loyalty can be fuelled with the quality of the e-service. One popular method of evaluating service quality on the Internet is IS SERVQUAL which includes reliability, responsiveness, assurance and empathy (Jiang, Klein, & Carr, 2002). The first three variables reliability, responsiveness and assurance measure related concepts to organizational and institutional trust, discussed in the following section. All four aspects of the service should be optimized at all times. Maintaining the minimal necessary service quality before an attack may build up insufficient e-loyalty to buffer the effect of the attack and result in a system switch. A service quality that exceeds user's expectations, will have a higher level of e-loyalty and more possibilities of retaining the user through the challenging period of an RW attack. Therefore, given the repeated RW attacks, organizations need to be prepared both on a technical level and in their ability to absorb the negative effect on users. A loyal user can use the same online service throughout a lifetime showing a degree of tolerance and forgiveness. If an incident, or series of incidents, causes the user to switch to an alternative online or offline service, the user can show lifelong loyalty to their new choice of system. Inertia can start building in favour of the new system, once it is adopted, and it will start to gain the advantages of an incumbent system. The previous incumbent system will now be an unappealing alternative, saddled with the dissatisfaction that led to the switch. There is, therefore, a lot at stake.

*Institutional Trust and Organizational Trust*

Trust has been proven to have an effect in human interaction particularly when exchanging value online such as when a user purchases a train ticket. Trust has also been found to have a significant influence on the consumer's e-loyalty towards a merchant (Carter et al., 2014). Trust can be distinguished between the organization the user is in contact with that provides the product or service and the institutions that support this exchange of value (McKnight & Chervany, 2002). These institutions include the Internet, the platform such the Google or Microsoft ecosystem, watchdogs and regulators. The organizations and institutions support a status quo of trust for many users that encourages them to engage in e-commerce. An example of how institutions influence organizations' behaviour online are the investigations into online vendors that have their data compromised. The investigations check if the compromised vendor's information systems and processes were suitable and met legal requirements. This is an example of how responsible behavior on the Internet is encouraged and enforced. A second example of the role of institutions are the cases when government regulators encouraged technology companies to provide updates to vulnerable systems at lower prices, to reduce the risk (FRPT Research, 2016).

*Information Privacy*

For a user to engage in many activities on the Internet, particularly acquiring e-services and products, some personal information must be shared. Over time most users have accepted that a level of personal information disclosure is necessary with certain vendors, platforms and institutions when exchanging value on the Internet (Chen, Zarifis, & Kroenung, 2017). This information can include a name, address and banking details. Some record of online activities including browsing and purchase history, may also be stored. This can be considered as an informal and formal agreement between the user and the Internet based organizations about how personal information is stored, shared, exploited and protected from threats (Conger, 2008). The formal part of the agreement can include the corporate privacy policies and legal framework. The user calculates the perceived control over the situation, the perceived risk and perceived intrusion from sharing this information (Xu, Dinev, Smith, & Hart, 2008). The user may only have a perception of how the personal information is used and may not be fully aware of the extent of the use. This perception can be informed in several ways such as targeted adverts. If the user purchased a ticket to Paris and they start receiving adverts for hotels in Paris, they may start to think about which organization shared their personal purchasing information. Given that RW attacks compromise the security of the organization targeted and encrypt data, personal information may also be accessed. If the user's personal information was accessed for it to be encrypted, was it also stolen? Would the organization know if it was also stolen? Would the organization inform them? These questions around privacy increase the perceived information asymmetry (Ba & Pavlou, 2002) between the user and the organization they are engaging with. Therefore, in the aftermath of an attack the user may reconsider the extent to which they are willing to share personal information.

### Research Model

The research model is based on the theoretical background covered in the previous section. The model presented in figure 2 identifies three variables that encapsulate the total effect of the RW attack and four variables that form the e-commerce status quo. The variables of the RW attack negatively influence the four variables of the status quo. The four variables of the status quo increase the resistance of the user to switching system (Polites & Karahanna, 2012). The proposed model therefore evaluates whether the status quo from the user's perspective is strong enough to absorb the repeated, extended RW attacks or if the attacks will cause an intention to switch systems. The RW attacks must be evaluated by taking into account the longitudinal and cumulative effect. This effect can be separated into three dimensions: Firstly, the specific effect on the user. This refers to the specific inconvenience the user endured such as a delay in booking a train ticket. The RW effect is hypothesized to negatively influence the four variables that form the e-commerce status quo:

**H1:** The ransomware attack impact on the user, will have a negative effect on the inertia towards switching system.
**H2:** The ransomware attack impact on the user, will have a negative effect on the trust towards institutions that support e-commerce.
**H3:** The ransomware attack impact on the user, will have a negative effect on trust towards the organization engaged with directly for the exchange of value.
**H4:** The ransomware attack impact on the user, will have a negative effect on the willingness to share personal information.

The second constituent part of the attack is the RW duration. It is hypothesized that the longer the duration is, the greater the negative impact on the constituent variables of the e-commerce status quo. For example, an attack that causes a core service failure for under one hour, over one hour but less than 24 hours, and over 24 hours will cause a progressively greater effect on the user's beliefs.

**H5:** The ransomware attack duration will have a negative effect on the user's inertia towards switching system.
**H6:** The ransomware attack duration will have a negative effect on the user's trust towards institutions that support e-commerce.
**H7:** The ransomware attack duration will have a negative effect on the user's trust towards the organization engaged with directly for the exchange of value.
**H8:** The ransomware attack duration will have a negative effect on the user's willingness to share personal information on the Internet.

The third and final constituent part of the attack is the RW repetition. It is hypothesized that the more times these attacks happen, the greater the cumulative negative impact on the constituent variables of the e-commerce status quo.

**H9:** The number of ransomware attacks will have a negative effect on the user's inertia towards switching system.

**H10:** The number of ransomware attacks will have a negative effect on the user's trust towards institutions that support e-commerce.

**H11:** The number of ransomware attacks will have a negative effect on the user's trust towards the organization they engage with directly for the exchange of value.

**H12:** The number of ransomware attacks will have a negative effect on the user's willingness to share personal information on the Internet.

The e-commerce status quo is formed by four constituent parts. These are inertia, institutional trust, organizational trust and information privacy. Inertia is created by the habit of using an incumbent system, the perceived transition costs and a psychological commitment due to perceived sunk costs and the appeal of the alternatives (Ghazali et al., 2016; Polites & Karahanna, 2012). The higher the level of the inertia, the harder it is for a critical incident to make a user switch system:

**H13:** The user's inertia will have a positive effect on the resistance towards switching system.

Institutional trust is shaped by cumulative positive experiences. Positive experiences include competent and effective related institutions including the government, regulators, Internet providers and software vendors. Additionally, structural assurance is indicated by the Internet environment being ordered and normal. Structural assurance is also supported by the availability of regulations and legal remedy on the Internet. Furthermore, institutional trust is shaped by the institutions' response to the attack. The initial response may stop short of solving the problem but it can be organized, coordinated and reassuring (Aliakbarlou, Wilkinson, Costello, & Jang, 2017). The final response must offer an effective and convenient solution. In addition to institutional trust the user must also trust the organization that value is exchanged with directly, such as a vendor. The user's belief in the competence, benevolence, integrity of the vendor builds this trust (McKnight & Chervany, 2002).

**H14:** The user's trust towards institutions that support e-commerce will have a positive effect on resistance towards switching system.

**H15:** The user's trust towards the organization engaged with directly for the exchange of value will have a positive effect on resistance towards switching system.

The fourth and last constituent part is the explicit and tacit contract between the user and the online organization on their personal information privacy. Perceived information privacy is formed by the existence of formal policies and procedures such as the perceived effectiveness of privacy statement, the perceived reasonableness of the data requested, the perceived reasonableness of the use and sharing of the information (Conger, 2008).

**H16:** The user's belief that personal information will be kept private, has a positive effect on resistance towards switching system.

## METHODOLOGY

The quantitative analysis applied Partial Least Squares Structural Equation Modelling (PLS-SEM) using the ADANCO software. As the model is moderately complex the analysis is in two stages. The measurement model is evaluated first and then the structural model. The measurement model evaluates the eight latent variables with three measured variables for each one. The structural model evaluates the hypothesized relationship between the eight latent variables.

**Figure 2. Research model: The impact of ransomware attacks on the user's intentions**

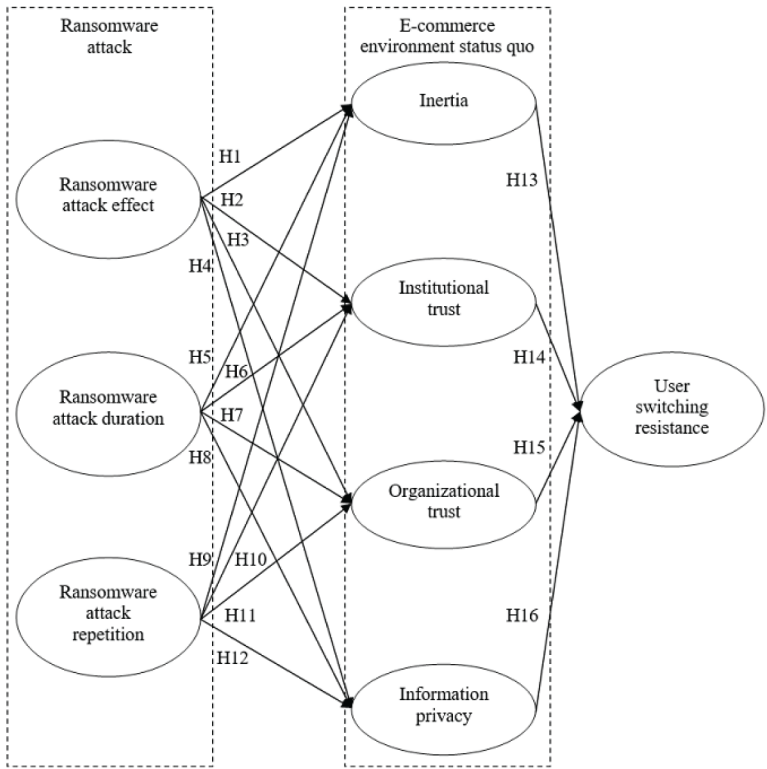

For the operationalization of the research variables a number of validated scales were utilized including scales for habit (Polites & Karahanna, 2012), trust (McKnight, Choudhury, & Kacmar, 2002), technology adoption (Venkatesh, Morris, Davis, & Davis, 2003) and information privacy (Hui, Teo, & Sang-Yong, 2007; Liang, Xue, Laosethakul, & Lloyd, 2005). The scales for RW attacks were developed by this research.

The sample was collected by the online survey tool SoSci Survey (www.soscisurvey.de) from the general population. There was a requirement that participants know what the RW attacks are and that they have experienced them either directly or indirectly. The necessary sample size for a significance level of 1% and a minimum $R^2$ of 0.10 was calculated to be 191 (Hair, Hult, Ringle, & Sarstedt, 2014). An alternative method using the G*Power software with an effect size of 0.5, a power of 0.95 and 8 degrees of freedom suggested a minimum sample size of 91. The survey was completed by 243 participants from which 204 are considered sufficiently complete and valid.

Table 1. Demographic information of the participants

|  |  | **Frequency** | **Percentage** |
|---|---|---|---|
| Gender | Female | 91 | 42 |
|  | Male | 118 | 58 |
| Age | Under 18 | 8 | 4 |
|  | 18-24 | 95 | 47 |
|  | 25-39 | 64 | 31 |
|  | 40-59 | 37 | 18 |
|  | 60 or older | 8 | 4 |
| Education level | Without educational level | 6 | 3 |
|  | High school | 103 | 51 |
|  | Undergraduate university degree | 56 | 28 |
|  | Post-graduate university degree | 39 | 19 |
| Income (in British Pounds per month) | No regular income | 17 | 8 |
|  | 400-1200 | 43 | 21 |
|  | 1201-3000 | 50 | 25 |
|  | 3001-5000 | 86 | 42 |
|  | > 5000 | 8 | 4 |

## ANALYSIS AND RESULTS

The PLS-SEM analysis is presented in two stages starting with the measurement model and then the structural model. The measurement model evaluates how strong the relationship is between the observed and latent variables. The structural model evaluates how strong the relationship is between the latent variables.

### Evaluating the Measurement Model

The measurement model is evaluated in a number of ways including the Fornell-Larcker Criterion presented in table 2 and the factor loading, Construct Reliability (CR) and Average Variance Extracted (AVE) presented in table 3. When the initial research model was tested the Fornell-Lacker criterion showed a strong correlation between two sets of variables. The first pair with a high correlation were ransomware attack effect and duration. The second pair with a high correlation were institutional trust and organizational trust. Therefore, the two pairs were merged leaving the model with six variables. Variables with a high correlation can be merged if this is compatible with the logic of the model and can also be supported theoretically from the literature (Hair et al., 2014). Merging the two dimensions of trust and the two dimensions of RW attacks meets these criteria. The Fornell-Lacker criterion measures for the new version of the model presented in table 2 did not have problematic correlations. The loadings of the observed variables that were reflective constructs, exceeded the required level of 0.70 (Hair et al., 2014). Construct Reliability (CR), calculated by the Cronbach's Alpha is over the necessary 0.70. This is an indication that the items that form the latent variable are sufficiently related. Average Variance Extracted (AVE) is above the threshold of 0.50 (Hair et al., 2014). The AVE indicates that the items explain the latent variable and there is enough convergent validity between them.

Table 2. Fornell-Larcker Criterion

| Construct | RAE | RAR | Inertia | Trust | IP | USW |
|---|---|---|---|---|---|---|
| RW attack effect (RAE) | 0.7207 | | | | | |
| RW attack repetit. (RAR) | 0.5138 | 0.8179 | | | | |
| Inertia | 0.1609 | 0.0881 | 0.7558 | | | |
| Trust | 0.1060 | 0.0843 | 0.7454 | 0.7305 | | |
| Information privacy (IP) | 0.0519 | 0.1230 | 0.6079 | 0.6341 | 0.7564 | |
| User switch resist. (USW) | 0.0729 | 0.0576 | 0.7583 | 0.8054 | 0.6354 | 0.8504 |

## Evaluating the Structural Model

The structural model is evaluated in several ways that are presented in table 4 and figure 3. The coefficient of determination $R^2$ is above 0.1 for Inertia, Trust and Information Privacy (IP) that can be considered weak and above 0.8 for User Switching Resistance (USR) which is substantial (Chin,

Table 3. Results of the measurement model

| Scale/Item | Loadings | CR | AVE |
|---|---|---|---|
| Ransomware attack effect | | 0.8063 | 0.7207 |
| RAE01 | 0.7967 | | |
| RAE02 | 0.8657 | | |
| RAE03 | 0.8820 | | |
| Ransomware attack repetition | | 0.8893 | 0.8179 |
| RAR01 | 0.9241 | | |
| RAR02 | 0.8604 | | |
| RAR03 | 0.9270 | | |
| Inertia | | 0.8382 | 0.7558 |
| I01 | 0.8989 | | |
| I02 | 0.8509 | | |
| I03 | 0.8574 | | |
| Trust (Organ. and Instit.) | | 0.8153 | 0.7305 |
| T01 | 0.8518 | | |
| T02 | 0.8309 | | |
| T03 | 0.8806 | | |
| Information privacy | | 0.8388 | 0.7564 |
| IP01 | 0.8637 | | |
| IP02 | 0.8520 | | |
| IP03 | 0.8929 | | |
| User switching resistance | | 0.8242 | 0.8504 |
| USR01 | 0.9235 | | |
| USR02 | 0.9208 | | |

1998). This suggests a weak explanatory power of the exogenous variables. The effect size is large for the path from Trust to USR as it is above 0.35 and moderate for Inertia to User Switching Resistance (USR) as it is above 0.15. The other paths are below 0.15 and therefore weak (Chin, 1998). Based on the final model there is support for seven hypotheses from the initial sixteen. While this could be considered low if it was a confirmatory analysis it is acceptable as this is an exploratory analysis. The findings validate the use of PLS-SEM in this research over Covariance Based SEM which is better for confirmatory analysis of more mature models.

Table 4. Results of the structural model

|  | Path coefficient | Standard error | t-value | p-value (2-sided) | Cohen's $f^2$ |
|---|---|---|---|---|---|
| RAE -> Inertia | 0.3875 | 0.0927 | 4.1826 | 0.0000 | 0.0870 |
| RAE -> Trust | 0.2417 | 0.1072 | 2.2543 | 0.0244 | 0.0320 |
| RAE -> IP | -0.0487 | 0.1040 | -0.4683 | 0.6397 | 0.0013 |
| RAR -> Inertia | 0.0190 | 0.0985 | 0.1926 | 0.8473 | 0.0002 |
| RAR -> Trust | 0.1171 | 0.0946 | 1.2378 | 0.2161 | 0.0075 |
| RAR -> IP | 0.3856 | 0.1009 | 3.8213 | 0.0001 | 0.0826 |
| Inertia -> USR | 0.3247 | 0.0550 | 5.8993 | 0.0000 | 0.1608 |
| Trust -> USR | 0.5026 | 0.0643 | 7.8179 | 0.0000 | 0.3596 |
| IP -> USR | 0.1437 | 0.0431 | 3.3361 | 0.0009 | 0.0453 |

Figure 3. Validated model: The impact of ransomware attacks on the user's intentions

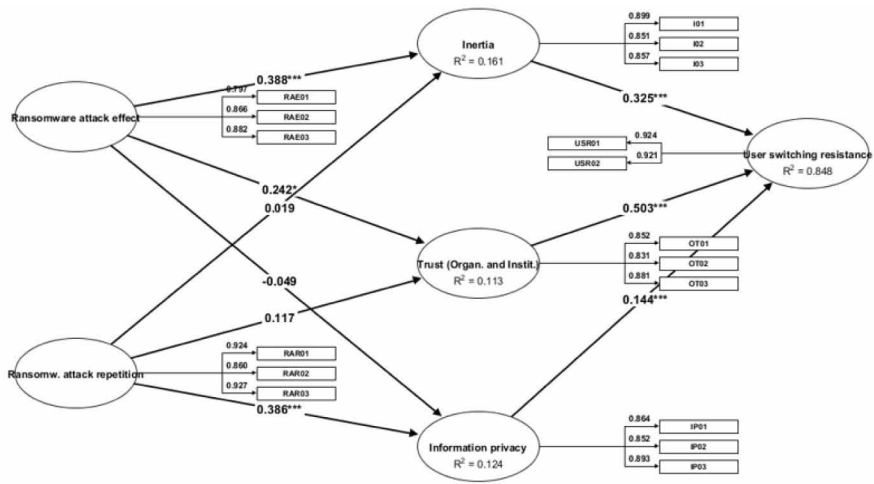

## DISCUSSION

### Theoretical Implications

The results of the analysis improve our understanding of the impact of RW attacks on the user and their intention to switch system. Trust in the organization and institutional trust have been evaluated separately in some research (McKnight & Chervany, 2002) while other research combines them (McKnight, Carter, Thatcher, & Clay, 2011). While this research distinguished between them in the initial model, the data showed that they are closely related from the user's perspective in this context. While there is support from the data for both Ransomware Attack Effect (RAE) and duration the data also shows that from the user's perspective, these are very closely related and to some degree indistinguishable. Therefore, it is recommended that future research use a variable covering both effect and duration together.

User inertia or e-loyalty (Carter et al., 2014; Polites & Karahanna, 2012) and personal information privacy concern were supported within the model. The path from trust to user switching resistance is stronger than the path from information privacy.

### Practical Implications

Based on the findings of this research a new strategy to reduce the effectiveness of an attack can be made: As the repetition of a RW attack influences the user, it is important for the organization to respond quickly. The response should have two parts: First, the attack should be stopped, and the service should be re-established. As the repetition of the attacks influences the user, the organization should not pay the ransom because this encourages further attacks. Secondly the three variables that support habit should be reinforced. Information privacy concerns should be reduced, trust must be built, and inertia should be strengthened. If this is not achieved, then the user may search for alternatives.

The attack should also be seen as part of the relationship with the user. Therefore, there should be a plan in place before an attack on how to interact with the user. The user may be required to take some steps during the RW attack such as change their password, re-enter data or update their security. It has been shown that users can adapt both positively and negatively to such requests (Chenoweth, Gattiker, & Corral, 2019) so it is important to have a strong relationship that will encourage a positive response from the user.

Lastly, organizations and their executives should show that they followed institutions guidance and all legal requirements so that they benefit from institutional trust. As users trust certain institutions, by associating with them this trust is transferred. In this way an impression of situational normality is also supported. Situational normality is conducive to trust. This research shows that repeated attacks can change habits so this provides further encouragement for executives, in addition to their legal requirements (Chatterjee, 2019), to take the recommended steps for Cybersecurity preparedness before, during and after a security breach.

## CONCLUSION

This research developed a model to evaluate whether a user would resist switching systems and stay within the e-commerce status quo despite extended and repeated RW attacks. The implications of this research are both theoretic and practical. A review of IS literature such as the AIS Electronic Library and MIS Quarterly, suggests that RW has received insufficient attention so far. Therefore, the first theoretic contribution is highlighting the importance of this issue to IS theory. The second, more specific, theoretic contribution is linking the RW literature to existing use of information systems. Furthermore, the theoretic understanding of inertia and status quo bias is expanded. The third theoretic contribution is exploring and developing a model on the impact of ransomware attacks on the user's witching intentions. After merging two couples of variables because of their strong correlation the final model with six latent variables is supported. Ransomware attack effect and repetition influence inertia, trust and information privacy. The last three variables influence user switching resistance. It is useful to identify which aspects of the RA attack have the most significant effect on the e-commerce status quo bias.

There are three practical implications. Firstly, these RW attacks rely on the disruption and fear they cause to achieve their goal of extracting a ransom. By better understanding the impact on the user, it may be possible to have a new strategy to reduce the effectiveness of the attack. Secondly, all online organizations that are potential victims can develop a number of processes to handle these attacks, going beyond the technical dimension. By better understanding the impact on the user, processes can be created to manage such disasters. Additionally, organizations such as online vendors can develop a buffer of goodwill that can absorb the negative impact on the user from an attack and stop them reaching the tipping point where they decide to switch system. Lastly, institutions can understand the mechanisms of a RW attack better and improve their approach both in preventing and resolving them.

Future research can evaluate the validity of the model with new RW attacks and explore possibilities of extending it to cover other security breaches or other events that have a large impact on the user.

*Alex Zarifis is passionate about researching, teaching, and practicing management in its many facets. He has taught in higher education for over ten years at universities including the University of Manchester, University of Liverpool, and the University of Mannheim. His research is in the areas of information systems and management. Dr Alex first degree is a BSc in Management with Information Systems from the University of Leeds. His second an MSc in Business Information Technology and a PhD in Business Administration are both from the University of Manchester. The University of Manchester PhD in Business Administration is ranked 1st in the world according to the Financial Times.*

*Xusen Cheng is a Full Professor in the School of Information at Renmin University of China in Beijing. He obtained his PhD degree from the University of Manchester, UK. His research is in the areas of information systems and management particularly focusing on online collaboration, global teams, the sharing economy, e-commerce, and e-learning.*

*Uchitha Jayawickrama is a Lecturer in Information Systems (which is equivalent to Assistant Professor) at the Information Management Group, School of Business and Economics, Loughborough University, UK. He has research, teaching and industry experience in the field of information systems. He is a Senior Fellow of Higher Education Academy (SFHEA), UK. Before joining Loughborough University, he was a Senior Lecturer and a Course Leader at Staffordshire University, UK. Dr Jayawickrama has published his research in various renowned conferences, books and journals (20+ peer-reviewed outputs) and has won 2 best paper awards in 2 international conferences. He is a reviewer for several journals and international conferences. He has editorial experience in various journals.*

*Simone Corsi joined Loughborough University from the University of Manchester, where he managed the Manchester China Institute and was Associate Member at the Manchester Institute for Innovation Research. Simone conducts research in the field of innovation studies. He has a research background in R&D and Innovation Management and has been looking into global innovation and reverse innovation dynamics, with a focus on China. More recently he has developed an interest in international R&D alliances and university-industry collaboration. Prior to joining the University of Manchester, Simone was Research Fellow at the Lancaster University Management School, where he was also Programme Manager for the Lancaster China Catalyst Programme, a Lancaster University project which aimed at creating and supporting R&D partnerships between UK and Chinese organisations.*